\documentclass[manuscript,screen,acmsmall]{acmart}

\AtBeginDocument{%
  \providecommand\BibTeX{{%
    \normalfont B\kern-0.5em{\scshape i\kern-0.25em b}\kern-0.8em\TeX}}}

\usepackage{url}
\usepackage{enumerate}
\usepackage{float}
\usepackage{longtable}
\usepackage{xltabular}
\usepackage{graphicx}
\usepackage{rotating}
\usepackage[utf8]{inputenc}
\def\markup{0}  
\if\markup 1
\usepackage{soul}
\usepackage[titletoc]{appendix}
\newcommand{\rv}[1]{{\leavevmode\color{blue}#1}}
\else
\newcommand{\rv}[1]{#1}
\newcommand{\st}[1]{}
\fi

\usepackage{graphicx}
\usepackage{xcolor,stfloats}
\usepackage{graphicx}
\usepackage{tabularx}
\usepackage{float}
\usepackage[section]{placeins}

\title[Practices and Challenges of Online Love-seeking Among Deaf or Hard of Hearing People: A Case Study in China]{Practices and Challenges of Online Love-seeking Among Deaf or Hard of Hearing People: A Case Study in China}

\author{Beiyan Cao}

\affiliation{
  \institution{The Hong Kong University of Science and Technology}
}
\email{beiyan.cao@connect.ust.hk}

\author{Changyang He}
\affiliation{
  \institution{The Hong Kong University of Science and Technology}
}
\email{cheai@cse.ust.hk}

\author{Jingling Zhang}
\affiliation{
  \institution{The Hong Kong University of Science and Technology (Guangzhou)}
}
\email{jzhang898@connect.hkust-gz.edu.cn} 

\author{Yuru Huang}
\affiliation{
  \institution{The Hong Kong University of Science and Technology (Guangzhou)}
}
\email{yhuang760@connect.hkust-gz.edu.cn}

\author{Muzhi Zhou}
\authornotemark[1]
\affiliation{%
  \institution{The Hong Kong University of Science and Technology (Guangzhou)}
}
\affiliation{
  \institution{The Hong Kong University of Science and Technology}
}
\email{mzzhou@hkust-gz.edu.cn}

\author{Mingming Fan}
\authornotemark[1]
\affiliation{
  \institution{The Hong Kong University of Science and Technology (Guangzhou)}
}
\affiliation{
  \institution{The Hong Kong University of Science and Technology}
}
\email{mingmingfan@ust.hk}

\begin{abstract}

People who are deaf or hard of hearing (DHH) in China are increasingly exploring online platforms to connect with potential partners. This research explores the online dating experiences of DHH communities in China, an area that has not been extensively researched. We interviewed sixteen participants who have varying levels of hearing ability and love-seeking statuses to understand how they manage their identities and communicate with potential partners online. We find that DHH individuals made great efforts to navigate the rich modality features to seek love online. Participants used both algorithm-based dating apps and community-based platforms like forums and WeChat to facilitate initial encounters through text-based functions that minimized the need for auditory interaction, thus fostering a more equitable starting point. Community-based platforms were found to facilitate more in-depth communication and excelled in fostering trust and authenticity, providing a more secure environment for genuine relationships. Design recommendations are proposed to enhance the accessibility and inclusiveness of online dating platforms for DHH individuals in China. This research sheds light on the benefits and challenges of online dating for DHH individuals in China and provides guidance for platform developers and researchers to enhance user experience in this area.

\end{abstract}



\begin{document}
\maketitle

\section{INTRODUCTION}
\label{INTRODUCTION}

China is home to over 20 million individuals who are deaf or hard of hearing (DHH), making it one of the largest populations with hearing disabilities in the world~\cite{ChinaDHH2006}. However, as a developing country, DHH individuals in China often face various forms of exclusion. This includes difficulties in securing employment opportunities~\cite{Wangxiaolan2016}, seeking legal support~\cite{ZhangRuixue2022}, and finding romantic partners~\cite{DHHDating, DisDate}. In contrast to some developed countries like the USA, where the DHH community has its own culture and identity~\cite{ASLCulture2011,deafp, Woods2020}, Chinese DHH individuals face challenges in establishing their own identity and pride. Many of them strive to be accepted as ``normal people" by mainstream society~\cite{Wangxiaolan2016, WangZiyan2022}, which further complicates their experiences, particularly in the realm of love-seeking~\cite{wenlin1993,DHHA30}. 

The growth of digital technologies within China's Internet market has provided online platforms that have the potential to enhance opportunities for DHH people in various domains, including education~\cite{Toofaninejad2017}, e-commerce~\cite{cao2023}, and legal support~\cite{ZhangRuixue2022}. For example, a recent news report highlighted that more DHH individuals are using online matchmaking services through livestreaming apps~\cite{DHHdatestream}. This suggests that online dating platforms are becoming increasingly popular among the DHH community, and there is a growing need to ensure that these platforms are accessible and inclusive for all users.

Minority groups encounter marginalization and exclusion on mainstream online dating platforms, which are not universally designed to accommodate diverse user needs. Individuals with autism~\cite{LoveAutism}, those living with human immunodeficiency virus (HIV) \cite{warner2019signal}, and wheelchair users~\cite{Santinele2022, Milbrodt2019} frequently face barriers, such as when deciding whether to disclose their identity. A particular issue for the DHH community is the reliance on audio features by users of mainstream dating platforms such as \textit{Hinge} \cite{Hinge} or \textit{Bumble} \cite{Bumble}. This reliance on auditory communication may potentially exclude DHH users, who are unable to access these features fully. It is not merely the presence of audio features that marginalizes DHH users but rather the lack of alternative, accessible communication methods that accommodate their needs. 


Studies about how DHH individuals date online and what challenges they meet remain very limited. There remains a notable gap in the research regarding the online dating experiences of DHH individuals. In Western societies, there have been well-established support systems and accessibility laws to provide a framework for inclusive communication and interaction~\cite{USAdeafLaw,AustraliaDeaf}. Nonetheless, recent studies have revealed a trend where, despite the availability of supportive legal frameworks, DHH individuals frequently hesitate to disclose their hearing disability when using online dating platforms~\cite{Zhu2023}. This hesitation underscores a discrepancy: theoretical supports do not fully translate into practical confidence and comfort for DHH individuals in online dating scenarios. 


Moreover, the DHH contexts differ widely across regions. For example, as one of the middle-income countries, China has just started to establish the disability supportive infrastructure~\cite{Richard2005, ZhangRuixue2022, Wangxiaolan2016}. There is also a lack of a widely accepted uniformed sign language system. For example, a small percentage, only 7.69\%, of DHH individuals in China can comprehend the sign language broadcasted on CCTV, the country's largest television network~\cite{cctvSign}. Instead, DHH individuals in China use regional informal sign languages~\cite{fischer_gong_2010,cao2023,Gabrielle2021}. The almost absence of social support and unified sign language make the online dating experiences of DHH individuals in China likely different from their counterparts in many high-income societies. In summary, the lack of understanding of the distinct experience of using online dating platforms for DHH individuals from China and the western countries may hinder the design of more accessible and inclusive online platforms. This can prevent DHH users from fully utilizing the potential benefits of online dating.

 
In this study, we aim to identify new opportunities to make these online love-seeking platforms more accessible globally by investigating the online love-seeking experiences of Chinese DHH individuals. We present our research questions (RQs) below:

\begin{itemize}
    \item RQ1: What motivates Chinese DHH individuals to seek love online?
    \item RQ2: What are Chinese DHH individuals' online love-seeking practices? Specifically, how do they disclose their DHH identity and interact on these platforms? 
    \item RQ3: What are the barriers that Chinese DHH individuals face when seeking love online? What coping strategies do they employ to overcome these barriers? 
\end{itemize}

We recruited 16 DHH individuals who used online platforms to form romantic connections in China and conducted semi-structured interviews. These interviews revealed that online platforms offer an inclusive starting point for conversations and relationships because of its limited dependency on auditory signals. Participants also adopted various timing and methods of disclosing their hearing status, tailoring their approach based on the stage of relationship development and their previous experiences. Our participants mainly utilized two types of online platforms: algorithm-based dating apps and community-based platforms, which provide opportunities to engage in equal communication and establish connections with DHH and hearing people. 

Yet, many DHH participants still faced unfavorable reactions, some even led to the move away from algorithm-based dating apps. Community-based platforms in China, with a likelihood presence of DHH individuals, greater community support, and human matchmakers following disability verification processes and having extensive social networks, have become more popular. These platforms have shown a higher success rate in matching individuals. In these specialized communities, the search for romantic connections flourished due to the sense of intimacy and trust created through shared experiences and authenticity, which resonated positively with the DHH community in China.

This study is one of the first to investigate the online dating experiences of Chinese DHH individuals. Our research provides a rich understanding of the strategies that Chinese DHH platform users employ to maximize their chances of success in locating a romantic partner. The current Chinese online platforms gain popularity with their rich modality features and large user groups; however, those mainstream, algorithm-based dating apps tend to pose challenges in identity management and interaction for DHH users. Moreover, we suggest design implications for addressing these challenges for DHH users. These findings have implications for designing more inclusive and accessible dating apps, both in China and globally.


\section{RELATED WORK AND BACKGROUND}
\label{RELATED WORK}
\subsection{Digital Technology Accessibility for DHH Individuals}

DHH individuals face numerous challenges in online communication. These individuals, especially Deaf signers, often have reading and writing skills that are lower than those of their peers, making it difficult for them to comprehend written language on social media platforms~\cite{mack2020social}. Moreover, with the global trend of multimodal communication on social platforms, the reduced reliance on text-based interactions highlights the importance of visual design for effective communication~\cite{alnfiai2017social, 10.1145/3311957.3359439}. However, without proper accommodations, such as video features with sign language translation and accessible text-to-speech interfaces, this multimodality design may widen the communication gap between DHH people and hearing people, creating barriers to effective communication~\cite{10.1145/3311957.3359439}.

Previous studies have investigated the barriers and needs of deaf individuals in online communication, such as social media interactions~\cite{cao2023}, educational platforms~\cite{Aljedaani2022}, and workplace communication sites\cite{Luft2000}, identifying user requirements for program design~\cite{chiu2010essential, alnfiai2017social}. For instance, Mrim Alnfiai and Srini Sampali examined 55 communication applications and found that only six were specifically designed for deaf users. They recommended that interpreters translate written text into high-quality video streams using sign language, developers incorporate accessibility features into available smartphone applications, and multiple methods (video streams, audio, and text) be used to present information~\cite{alnfiai2017social}. Moreover, MobiDeaf, a set of guidelines, has been developed to support developers in creating web and mobile applications for deaf users~\cite{schefer2018supporting}. For instance, the importance of simple interfaces and the necessity of clear visual or vibrating notifications have been highlighted to ensure that deaf users can effectively navigate and interact with the application.

To meet the needs of DHH individuals, more accessible information and communication technologies are being developed. Scholars have suggested various strategies, such as non-audible feedback~\cite{schefer2018supporting}, sign language assistance~\cite{alnfiai2017social}, and automatic speech recognition(ASR)~\cite{seita2022remotely}, with many works focus on improving the recognition accuracy of ASR~\cite{10.1145/3132525.3134798,seita2022remotely}. Besides accuracy issues, non-verbal elements (such as speech rate and volume) are also considered important information by DHH individuals~\cite{mcdonnell2021social}. To address this, three novel captioning models have been developed, which, in addition to text, depict prosody, emotion, and a combination of both~\cite{10.1145/3544548.3581511}. 


Despite these efforts to develop more accessible information and communication technologies, there is a noticeable absence of discussion regarding the more specific needs of DHH individuals when using online dating sites. Accessible information and communication are crucial in online dating, from assessing profiles to engaging in deeper interactions, yet how DHH individuals interact with these platforms and others during online love-seeking remains under-explored.


\subsection{Identity Management in Online Dating}
\label{section 2.2}

Identity management, which usually involves self-presentation, highly influences online dating experiences. Self-presentation or impression management refers to the way people present aspects of themselves through various aspects such as appearance in photos~\cite{10.1145/1357054.1357181}, demographics, values~\cite{10.1145/2851581.2892482}, personalities~\cite{10.1145/2851581.2892482}, voice~\cite{Shen2024}, and even friend list~\cite{10.1145/1316624.1316680} to appeal to others~\cite{goffman2016presentation}. Compared to traditional online dating applications where text-based communication dominates~\cite{finkel2012online}, recent online dating platforms have developed diverse modes of self-presentation, including photos, voice messages, and even videos~\cite{Shen2024}.


Online love-seeking platform users strategically disclose information and present an idealized version of themselves~\cite{fitzpatrick2018shut, whitty2008revealing, 10.1145/1240624.1240697}. For example, users may intentionally alter their voices when using a voice-based dating app, ~\cite{Shen2024}. In addition, online dating participants often misrepresent themselves, including physical appearance ~\cite{gibbs2006self}, age~\cite{article}, and marital status~\cite{article}. Therefore, online dating participants report that the widespread deception is the ``main perceived disadvantage of online dating"~\cite{article, zytko2014impression}. Consequently, before meeting in person, users often lack confidence in their impressions of their communication partners~\cite{zytko2014impression}. This anticipation of face-to-face interaction can influence their choices in self-presentation~\cite{walther1994anticipated}. As the perceived probability of future face-to-face interaction increases, people more closely monitor their disclosures~\cite{giles1979language} and are more likely to disclose authentic information~\cite{gibbs2006self}. Porter et al.\cite{Porter2017} explored the disability disclosure preferences of adults with and without disabilities who have dated online.

Recent research has expanded the understanding of self-presentation in online dating to marginalized groups. Fernandez et al.~\cite{Fernandez2019} explored how transgender individuals manage their self-presentation in online dating environments. This research highlights how transgender platform users balance the need for honesty with concerns about safety and acceptance. Similarly, Hardy et al.~\cite{Hardy2017} examined the self-presentation strategies of \textit{Grindr} users in rural areas, where the population size is small and the users' true identity is highly likely to be identified. They found that rural users often carefully curate their profiles and interactions to avoid unwanted attention and potential discrimination. 

Zytko et al.\cite{zytko2014impression} investigated the broader processes of impression management and formation in online dating systems and noted that users are often frustrated by the ambiguity of online interactions and the difficulty in accurately conveying their personality and intentions through profile information. This frustration can be even more pronounced for marginalized individuals who must navigate additional layers of identity management and disclosure. In a more recent work, Datey et al.~\cite{Datey2022} explored strategies for disclosing and detecting sexual interest. They found that users employ a range of subtle and overt cues to communicate their intentions, and the effectiveness of these strategies is influenced by the design of the dating platform. 

While existing studies offer insights into how marginalized groups handle self-presentation and impression management in online dating, there is a limited focus on DHH individuals. We have found only one study investigated 1,200 dating posts in \textit{AllDeaf}, which is an English online community and resource hub for people with hearing loss, to uncover identity disclosure patterns in dating~\cite{Zhu2023}. The study found that DHH individuals are reluctant to disclose their disability while openly discussing their relationship difficulties, especially in relationships between hearing and DHH individuals. However, the experiences of DHH individuals seeking love on mainstream dating platforms are rarely explored or emphasized.

\subsection{Cultural and Social Context for Chinese DHH People in Online Love-seeking}

Finding a partner is considered to be challenging for DHH individuals in China. The lack of social support towards disabled people in general may be the main reasons that contribute to the discrimination and inequity they face in various aspects of their daily lives~\cite{ChinaDHH2006,DisDate}. There is a shortage of specialized training programs for educators in deaf education~\cite{article, inbook}. In legal matters, DHH individuals sometimes struggle to obtain appropriate assistance when they are subjected to deception~\cite{ZhangRuixue2022}. In the workforce, DHH individuals still face limited employment options, workplace discrimination, and inequitable compensation~\cite{Wangxiaolan2016}.

The social culture, believing in ``compensating for disabilities'', further creates barriers for DHH individuals to feel proud or be respected in the society. This norm emphasizes rehabilitation with the belief that acquiring the ability to hear and speak equates to recovery, especially among DHH children and their parents~\cite{deafcureChina}. This approach contrasts with the celebration of deaf identity and culture observed in some Western countries, such as the US, where there is greater acceptance and recognition of deaf culture~\cite{ASLCulture2011,deafp, Woods2020}. The existing social attitudes and limited support systems make online dating platforms appealing to Chinese DHH individuals due to their large user base. Furthermore, this scenario can also present challenges in the online love-seeking experiences of Chinese DHH individuals because the stigma and discrimination toward this group are also extended online.

Moreover, the diversity and regional variations of Chinese Sign Language present challenges for Chinese DHH individuals in online love-seeking. This linguistic diversity hinders the formation of a cohesive DHH community and the establishment of a unified DHH-pride movement~\cite{ali2021,le2004,monaghan2003,perniss2008visible,pfau2016}, which are crucial for conveying social norms and fostering a sense of belonging and identity~\cite{Ahlin2023}. Furthermore, the stigma and discrimination they face offline may also extend to online interactions, complicating communication and relationship building. Despite growing opportunities in online love-seeking, limited research exists on how these factors may impact Chinese DHH individuals. This gap indicates a need for further investigation into their motivations, practices, and challenges in online dating in China. Our study aims to bridge this gap by investigating how Chinese DHH date online.

\section{METHOD}
\label{METHOD}
We conducted semi-structured interviews with sixteen Chinese DHH participants. All participants have used various Chinese mainstream online platforms (i.e., \textit{Tantan}, \textit{Momo}, \textit{Zhenai}, \textit{Qingtengzhilian}, \textit{Ergou}, \textit{MarryU}, \textit{Tashuo},\textit{Douban}, \textit{Tieba}, \textit{WeChat groups}, \textit{WeChat public accounts},\textit{WeChat private account}, etc.) in their quest for romantic partners. These interviews were complemented by a pre-interview survey designed to gather basic information such as age, sex, state of DHH (following the Chinese Practical Assessment Standard for Disabled Persons ~\cite{DHHstandard}), communication (speaking and hearing) state, current relationship status, past experiences and frequency of use online platforms. This will help us to tailor our interviews to be more effective as preliminary information is known. This study received ethical approval from the university.

\subsection{Procedure}
\label{Procedure}

In this study, we initially recruited participants by advertising on \textit{Xiaohongshu} (the little red book) \cite{xhs}, which is a rapidly growing social media platform in China, has garnered significant popularity by facilitating the sharing of meticulously curated short videos, photographs, and images. The advertisement included a QR code that interested people could scan to sign up. The selection criteria were individuals aged 18 and older who hold an officially issued DHH registration card or certificate from the Chinese government and have attempted to find romantic partners online. We provided the qualified participants with an interview consent form and a preliminary survey. In the survey, we asked for basic demographic information such as gender, age, DHH state, as well as their usage of online dating websites and the frequency of use (A sample is in Appendix.\ref{Questions}). 

Through posters on \textit{Xiaohongshu} and the snowball sampling method, we received 39 questionnaires filled out by DHH individuals who met our criteria for the interviews. Information of the final 16 participants for the interviews was summarized in Table.\ref{table:participants}. We used semi-structured interviews to better investigate the participants' experiences with online love-seeking. The interview was customized based on the participants' demographics and particular circumstances during interviews. For example, we aimed to delve deeper into the details of these situations, including the specific platforms used and the mode of communication (voice, text, etc.), if a participant was willing to share positive experiences from their online love-seeking journey. Conversely, we omitted questions for those who preferred not to discuss these topics.

The interviews were conducted online from March 21 to June 18, 2023. Participants decided their preferred interview format, including text messages, audio, video, or other modes. Three participants chose to conduct interviews via voice calls on \textit{WeChat} and others used text messages on \textit{WeChat}. Voice interviews lasted between 45 minutes to one hour, while text interviews were more time-consuming, with each participant's interview lasting from one and a half to three hours. Then, we meticulously recorded and transcribed all sessions for subsequent analysis. After the interview, we gave each participant 80 CNY\rv { (about 14 USD)} to compensate for their time.

\subsection{Data Analysis}
\label{Data Analysis}


Our data comprised audio recordings of online meetings and text messages from \textit{WeChat}. The audio recordings were transcribed into textual scripts using \textit{iFLYTEK}~\cite{iFlytek}, a popular Chinese speech-to-text mobile application. Five researchers from our team, all native Chinese speakers, reviewed the textual scripts to develop a first understanding of the online love-seeking experiences of DHH participants. 

To analyze the data, we began with an open coding approach to extract major concepts of DHH users' motivations, practices and challenges in online love-seeking. The open coding approach, as a key step of the grounded theory method~\cite{corbin2014basics}, allows codes to emerge from the interviews without predetermined categories, which sets the foundation for subsequent qualitative analysis. The process is as follows: Two researchers first independently familiarized themselves with the transcripts by thoroughly reading them to understand the context and content. Each coder generated initial codes that can summarize the ideas or concepts that emerged from the transcript. For example, when participants described the timing of revealing their DHH identities, we assigned these texts to the code ``Time to Disclose''. The two coders then compared and discussed their initial codes, developing a preliminary codebook with definitions and examples for each code. Discrepancies between the two coders were discussed in team meetings with the remaining three researchers, who helped resolve disagreements and refine the codes. Through rounds of team discussions, a consensus was reached. 

Further aggregation of the codes is guided by affinity diagramming~\cite{Muller2014} to organize and interpret the data further. Initial codes were written on sticky notes and grouped into clusters based on similarities to form higher-level groups and then into conceptual categories by the two coders together. Conflicts regarding organization of codes between the two coders were resolved in team meetings with the other three researchers. For example, ``Reasons to Disclose" and ``Time to disclosure" were grouped together into the ``Self-disclosure'' theme. This process is overseen by a senior researcher who specialized in HCI for socially marginalized groups. Our findings were eventually summarized into three high-level themes: \textit{Platform Empowerment}, \textit{Identity Management}, and \textit{Interaction Practices}. 

\subsection{Participants}
\label{Participants}

\begin{table}[htbp]

\caption{Summary of Chinese DHH participants interviewed. Among the 16 participants, 8 were female and 8 were male, aged from 22 to 39. DHH state \uppercase\expandafter{\romannumeral1}——Speech recognition rate \textless15\%; DHH state \uppercase\expandafter{\romannumeral2}——Speech recognition rate 15\%-30\%; DHH state \uppercase\expandafter{\romannumeral3}——Speech recognition rate 30\%-60\%; DHH state \uppercase\expandafter{\romannumeral4}——Speech recognition rate 61\%-70\%. This classification standard is promulgated by the Chinese government\cite{DHHstandard}.}

\scalebox{0.95}{
\begin{tabularx}{\textwidth}{X|X|X|p{0.6cm}|p{2.6cm}|p{2cm}|p{1.5cm}|p{1.5cm}|p{1.3cm}}
    \toprule
    ID    & Sex & Age   & DHH State & Communication State & Love-seeking State & Online Love-seeking History & Online Love-seeking Frequency & Dating Preference  \\ \hline
    P1    & M     & 23    & \uppercase\expandafter{\romannumeral2}     & CI User, Verbally Fluent & Developing  & 3 years & Once a week & non-DHH\\ \hline
    P2    & M     & 26    & \uppercase\expandafter{\romannumeral4}     & CI User, Verbally Fluent & Seeking & 1 year & Once every 2 days & non-DHH\\ \hline
    P3    & F     & 25    & \uppercase\expandafter{\romannumeral2}     & Hearing Aid User, Verbally Stutter & Seeking & 3 months & 6 months & non-DHH\\ \hline
    P4    & M     & 22    & \uppercase\expandafter{\romannumeral2}     & Hearing Aid User, Verbally Stutter & Seeking & 1 year & 2 hours everyday & no preference\\ \hline
    P5    & M     & 29    & \uppercase\expandafter{\romannumeral1}     & Hearing Aid User, Verbally Stutter & Seeking & 1 year & Everyday & non-DHH\\ \hline
    P6    & F     & 23    & \uppercase\expandafter{\romannumeral1}     & Hearing Aid \& Sign Language(SL) User & Seeking & 4 years & 30 mins everyday & non-DHH\\ \hline
    P7    & F     & 26    & \uppercase\expandafter{\romannumeral2}     & Hearing Aid User, Verbally Stutter & Developing  & 5 years & 4-5 days a week & non-DHH\\ \hline
    P8    & F     & 29    & \uppercase\expandafter{\romannumeral1}     & Hearing Aid User, Verbally Fluent & Developing  & 6 months & 3 days a week & non-DHH\\ \hline
    P9    & M     & 39    & \uppercase\expandafter{\romannumeral1}     & Verbally Fluent & Developing  & 1 year & Once every 2 weeks & DHH\\ \hline
    P10   & M     & 29    & \uppercase\expandafter{\romannumeral1}     & Hearing Aid User, Verbally Stutter & Seeking & 6 years & Everyday & no preference\\ \hline
    P11   & F     & 29    & \uppercase\expandafter{\romannumeral1}     & Hearing Aid \& SL User, Verbally Stutter  & Seeking  & 6 months & 2-3 days a week & DHH\\ \hline
    P12   & M     & 27    & \uppercase\expandafter{\romannumeral1}     & Hearing Aid User, Lip reader, Verbally Stutter  & Seeking & 7 months & Once a week & no preference\\ \hline
    P13   & M     & 35    & \uppercase\expandafter{\romannumeral1}     & CI \& SL User, Lip reader  & Married & 3 years & 3-4 days a week & no preference\\ \hline
    P14   & F     & 28    & \uppercase\expandafter{\romannumeral1}     & CI User, Lip reader, Verbally Fluent  & Developing & 3 years & Everyday & non-DHH\\ \hline
    P15   & F     & 30    & \uppercase\expandafter{\romannumeral1}     & CI User, Lip reader, Verbally Fluent  & Married & 13 years & Everyday & non-DHH\\ \hline
    P16   & F     & 26    & \uppercase\expandafter{\romannumeral1}     & Lip reader, Verbally Stutter  & Married & 1 year & Once a week & no preference\\ \hline

    \end{tabularx}%
}
\small
\centering
\Description[Summary of our DHH participants]{A table with 9 columns and 17 rows. Participants' Information is given, including ID, sex, age, DHH condition, love-seeking state, history, frequency, and preference for potential partners. There were eight females and eight males. eleven~(N=11) participants were in \uppercase\expandafter{\romannumeral1} DHH level (Speech recognition rate \textless15\%) , four~(N=4) participants were in \uppercase\expandafter{\romannumeral2} DHH level (Speech recognition rate 15\%-30\%) and one~(N=1) participants were in \uppercase\expandafter{\romannumeral4} DHH level (Speech recognition rate 61\%-70\%) . Participants were between 22 and 39 years old~($M=27.88, SD=4.36$) with online love-seeking history ranging from 3 months to 13 years ($M=32.88, SD=39.05$, unit=months). Nine participants wore hearing aids, and five participants wore Cochlear Implants (CI). Eight participants were still searching for partners online; five participants were developing their relationships online; and three participants were married to someone they met online.
We also asked participants whether they would prefer to find a partner who is also deaf or hard of hearing (DHH). Out of the total participants, nine preferred non-DHH partners, two preferred DHH partners, and five did not show a preference for either.}

\label{table:participants}
\end{table}

Table.\ref{table:participants} shows a summary of participants' information. Our participants (eight females and eight males) have various DHH conditions, love-seeking statuses, histories of online dating, and the frequency of using online platforms. Eleven participants were in DHH level \uppercase\expandafter{\romannumeral1} (Speech recognition rate \textless15\%) , four participants were in DHH level \uppercase\expandafter{\romannumeral2} (Speech recognition rate 15\%-30\%) , and one participant was in DHH level \uppercase\expandafter{\romannumeral4} (Speech recognition rate 61\%-70\%). This classification standard is promulgated by the Chinese government~\cite{DHHstandard}. Nine participants wore hearing aids, and five participants wore Cochlear Implants (CI), and the remaining participants did not use any hearing devices. 

Participants were between 22 and 39 years old~($M=27.88, SD=4.36$) with online love-seeking history ranging from three months to thirteen years ($M=32.88, SD=39.05$, unit=months). Eight participants were still searching for partners online; five participants were developing their relationships with someone they met online; and three participants were married to someone they met online. We also asked participants whether they would prefer to find a DHH partner. Nine of our participants preferred non-DHH partners, two preferred DHH partners, and five showed neutral preference.


\section{FINDINGS}
\label{Findings}


Our findings are grouped into three themes - \textit{Platform Empowerment}, \textit{Identity Management}, and \textit{Interaction Practices}. We began by presenting how online platforms attracted and facilitated DHH individuals to seek love online in Section \ref{findings: RQ1} (RQ1). In Section \ref{findings: Identity Management}, we provide a comprehensive analysis of how DHH participants navigate their identities when seeking romantic relationships online (RQ2). This section also explores the challenges that arise in relation to their identities (RQ3). Subsequently, in Section \ref{findings: Interaction}, we examine the communication strategies employed by DHH participants and the variety of community-based platforms they utilize (RQ2), along with the obstacles they encounter (RQ3).

\subsection{Platform Empowerment}\label{findings: RQ1}
Online platforms offer DHH individuals a convenient way to communicate through text. This form of communication provides more opportunities, especially at the beginning of a conversation or relationship, enabling DHH individuals to express themselves effectively through words and redirect focus from their DHH status. It creates a level playing field for DHH individuals to engage with others. P1 said \textit{``I find it easier to express myself through text online."}. P16 also said it was comfortable to use text instead of audio/video at the starting point of a relationship. P13 also stated,
    \begin{quote}
        \emph{``I always check her (online text) updates (now the woman is my wife) to see if I can find a suitable topic to talk about and identify the right starting point. I think I am really good at texting online. I mean, if she texts me for specific details, I won't directly reply to her; instead, I'll change the subject. Then, when she asks me again the next time, I'll share a small part, piquing her curiosity, and save the rest for our next conversation. "}
    \end{quote}

Moreover, online communication can help to overcome the difficulty in sound recognition that often exists in offline communication scenarios, especially during the initial dating events that usually happen in public areas. P1, who wore cochlear implants (CI), mentioned,  \textit{``I have avoided noisy environments knowing that my hearing works better in quieter settings."} P2, who also wore CI, added, \textit{``Offline meetings often involve noisy environments, making it difficult for me to hear clearly. This is why I prefer online interactions."} Likewise, P8 said, \textit{``Because the dining environment can be noisy at some shopping malls. The noisy environment in some shopping malls makes it difficult to gather verbal information, especially when it's crowded."} 


Another reason our participants used online platforms for love-seeking is that they were able to easily connect both within and outside of their communities. As P3 said, \textit{``My sense of existence or belonging can only be found online because it is tough for me to discover my other half in real life."}

In summary, text-based communication on these platforms has played a crucial role in overcoming the challenges faced by DHH participants in noisy offline environments. This mode of interaction has promoted fairer and more confident communication, assisting DHH users in initiating and cultivating relationships. Online platforms provide our participants with a simpler and more accessible means to connect and expand their social circles compared to offline settings.

\subsection{Identity Management}\label{findings: Identity Management}

A unique aspect of the DHH community is that their hearing ability is not immediately apparent during initial text-based interactions but becomes evident when meeting in person. Therefore, the timing and method of disclosing their DHH identity present an intriguing topic for exploration.

\subsubsection{\textbf{Self-Presentation through Profiles}}
 Our participants carefully curated their online profiles by sharing personal details, partner preferences, daily life, and hobbies to attract potential matches and simplify mutual selection. For instance, P7 and P8 shared detailed personal information and specific criteria for their ideal partners. 
 
\subsubsection{\textbf{Self Disclosure}}
\label{Section4.2.2}
All participants reported that they would disclose their DHH identity when searching for romantic partners online. In this section, we explore the reasons as well as the timing and techniques they employed to do so. Additionally, we report the responses they received after their disclosure.

\textbf{Reasons to Disclose.} Many participants explained that it was important to maintain honesty and a positive impression in front of their potential partners. This implies that the DHH identity would be revealed as soon as they met their dates offline, as P1, who was fluent in spoken language, reported, \textit{``I must inform the girl about my situation before meeting with her offline. I will feel deceitful if I keep hiding it from her, which will only make things worse."} P6 also said, \textit{``I don't want to hide it (DHH identity). I'd be cheating if I didn't say it at the beginning."} P10, who could use spoken language but with some stuttering, mentioned that his online date would eventually find out. Similarly, in P4's opinion, the DHH identity was a fact that could not be hidden.

Linked with this feeling of dishonesty, some of our participants mentioned the disclosure as a way to set expectations. P3 elaborated, \textit{``they will not have too high expectations of me before we physically meet offline."} P12 explained there might be misunderstandings if deception occurred initially.

Identity disclosure also works as a filter to select preferred dates, as P3 suggested, \textit{``If I am honest about my hearing condition from the beginning, those who can accept me will accept me."} P7, who disclosed her post-lingual deaf identity directly on her dating app's profile, stated that it's time-saving and convenient to select potential dates. She said,
    \begin{quote}
    \emph{``It saves trouble, otherwise it would be too troublesome to explain every time people ask. Also, disclosing my DHH condition on my profile is a two-way selection process. Only people who can accept my condition will come and talk to me on dating apps."} 
    \end{quote}
    
Additionally, P8 mentioned that she would disclose her DHH condition much earlier than usual when she wanted to end the conversation.

\textbf{Time to Disclose.} As aforementioned, the timing of when our participants disclosed their DHH status depended on their feelings about the potential partner and the progress of the relationship. We have gathered more details about the timing of disclosure, whether it occurred in the first contact or after lengthy communication.

A few of our participants disclosed directly on their online dating profiles. P7 explained, \textit{``It saves time"}. P14 disclosed her hearing condition on the original post of her dating thread on \textit{Douban}. Similarly, P11 also directly posted her hearing condition on \textit{Tieba}, a popular Chinese Forum where there were many disabled people.

13 out of the 16 individuals we interviewed opted not to disclose their DHH status on the profiles. P2, P3, P4, and P10 all expressed concerns that revealing their hearing loss to the public might lead to a very low contact rate, as expressed by P2, \textit{"I probably could not even start a conversation! "} P3 shared a sense of vulnerability, admitting she didn't feel resilient enough to be so candid early on. P6 expressed the importance of being treated equally, explaining, 
    \begin{quote}
            \emph{``I have just avoided being treated differently to find an opportunity for myself online. If I disclose that I have hearing loss at the outset, the other person will always be sympathetic during our subsequent conversation, and our level of communication will not be fair."}
    \end{quote}

Others disclosed their identity at different time points once the conversation had started. Some disclosed at the beginning of a conversation, as explained by P3, 
    \begin{quote}
        \emph{``I think it would be better to make it (DHH identity) clear at the beginning. Otherwise, both (me and my potential dates) will be satisfied when we text online, but they will be disappointed when we meet. This is because they have high expectations of me. The higher the expectations for me, the higher the disappointment could be."}
    \end{quote}
P5, shared similar thoughts, \textit{``When I am contacting someone or someone else adding me and taking the initiative to contact me, I will always be honest. When I chat with them, I will say it (my DHH identity) at the beginning of the conversation"} 

Nine participants decided to reveal their DHH identities as they got to know more about their dates. P1 mentioned that he would disclose after having texted a lot with his dates. P2 expressed, \textit{``Get to know her well and then tell her (DHH identity) at the right time. If she doesn't want to talk to me anymore, then I will delete her account."} Similarly, P3, P4, and P11 shared a similar approach. P6 added that, in addition to good communication quality, she also needed her dates to act polite, \textit{``Then I will consider telling my dates my hearing condition."} P9, P10, and P12 chose to disclose when they received positive signals from their dates while chatting online. As P12 explained, \textit{``That is, she thinks that I have a good character, a good figure, and we are having a very speculative conversation. In her eyes, I may be suitable for her, and then I would consider telling my date about my hearing condition."} 

\textbf{Dynamic Disclosure Strategies.} Several participants employed a dynamic strategy to disclose their DHH identity based on their learning from past experiences. For example, P3, who now revealed her DHH identity early, previously did not do so. She mentioned, \textit{``In the past, I kept my DHH identity a secret and only mentioned it when I and my dates first met offline. They (my dates) initially treated me with a lot of tolerance, but then they stopped talking to me anymore."} This experience led her to change her approach and disclose her DHH identity earlier in the dating process. Similarly, P11 changed her disclosure pattern from hiding to posting directly at the beginning because of efficiency.

In contrast, P4 delayed his disclosure. He explained, \textit{``I tried to indicate that I was DHH at the beginning, but in the end, no one contacted me or responded to my messages. So I learned from experience."} Similar to him, P8, who could use spoken language without any stuttering, said  \textit{``I once made a voice call to tell him (my identity). His voice sounded nice at that time. But he stopped sending me any text the next day. After that, I never mentioned my hearing status at the beginning."}   

Others revealed their DHH identity more tactically. P4, who wore hearing aids and could speak a little, chose to give some hint first by telling others that his work needed little talking. \textit{``I'll lead them carefully before finally stating that I have hearing loss."} P8, with no difficulty speaking or hearing with hearing aids, used different disclosing strategies towards various persons on dating apps. She pushed back the disclosure timing as much as feasible if she valued her date and was satisfied with him. On the contrary, if she had no interest in him, she would inform him about her hearing identity immediately, 
    \begin{quote}
        \emph{``I simply use this (disclosure) to convince him to give up. If I can't convince him to quit, he actually demonstrated his sincerity to me. Then we can stay in touch." }
    \end{quote}        

\textbf{Reactions After Disclosure.} Our participants received various reactions after they disclosed their DHH identities. 

Most reactions are rejections from hearing users but exceptions do exist. For example, P1 told us that his disability seems to inspire maternal instincts in some girls. He said,
    \begin{quote}
        \emph{``Some girls who had never dated a DHH person before are also really enthusiastic about the situation and wanted to give it a try."}
    \end{quote}    

Some comments were considered to be friendly. P6 gave some examples, \textit{``A man told me a deaf person is the kind of angel who communicates differently."} 

Though some participants were fortunate to receive some positive words after identity disclosure, the majority experienced varying kinds of challenges, which we present in the following chapters.

\subsubsection{\textbf{Challenges of DHH Identity Disclosure}}
We discovered three types of challenges regarding our participants' identity disclosure: social exclusion, misconception, and a lack of community support.

\textbf{Social Exclusion.} A common sentiment among our participants was a sense of isolation from the Chinese mainstream dating culture, which is defined by hearing people. They expressed concerns about possible rejection due to their hearing status and the way they were perceived by others on online platforms.

P3 articulated her concern about the hearing user-based algorithm-based dating apps, \textit{``Given that the majority of these dating app users can hear, I'm always a bit worried about how they might react to my hearing condition."} P12 spoke of deeper fears regarding family acceptance and genetic concerns, \textit{``I often face questions from women about the hereditary nature of deafness. They worry about the likelihood of having a DHH child (with a DHH individual). Even if my partner is understanding, we still have to consider whether her family can accept my hearing condition."} 

Our participants frequently encountered both explicit and subtle forms of rejection after revealing their DHH status, which often resulted in feelings of confusion, self-questioning, and heightened anxiety. These adverse reactions contributed to a sense of alienation on those algorithm-based dating apps. P4 shared the feeling of disappointment with the experience of repeated rejection,
    \begin{quote}
        \emph{``Whenever I mentioned my hearing condition to people I met through dating apps, their initial surprise would typically be followed by a polite refusal. They would express their uncertainty about engaging with someone with my condition, while still offering good wishes. At times, I wouldn't receive any direct response, which was disorienting, though I've gradually grown used to such outcomes."}
    \end{quote}
             
The absence of a clear rejection did not mitigate the feeling of exclusion for our participants. P5 recounted the change in dynamics after disclosing their deafness, 
            
    \begin{quote}
        \emph{``Upon learning of my deafness, I noticed a decline in the interest of the people I was talking to. Their initial enthusiasm faded, replaced by an apparent lack of engagement. This pattern has led me to be wary of dating apps and to think twice before recommending them to others in the DHH community, to avoid causing them similar distress. Video calls would often end abruptly once my hearing status became known. Even if I greeted them or tried to continue the conversation after explaining that I couldn't hear, the call was likely to be cut short. This response might be attributed to people's unfamiliarity with interacting with someone who is DHH."}
    \end{quote}
    
P9 also experienced ceased conversation once he disclosed his hearing status. Participants often encounter various excuses for stopping the conversation without directly referencing the fact that the other person is DHH. P3 shared insights into the indirect ways people would disengage,
    \begin{quote}
        \emph{``Some people choose to withdraw from the conversation after discovering my hearing condition, but rather than stating this outright, they provide different explanations. A common one is the challenge of distance, with remarks like, 'I'm not looking for a long-distance relationship, and we're too far apart.'"}
    \end{quote}  
    
P11 also reported facing indirect rejections, noticing a pattern where conversations would dwindle or stop entirely after the revelation of her hearing status.
             
Our DHH participants reported experiencing heightened feelings of isolation during their search for romantic connections online. P9 shared his proactive approach and the varied responses they encountered, \textit{``I took the initiative to tell others that my hearing is not good. While some respond with kindness, others react with hostility, with comments that can be hurtful and discriminatory. For instance, ``you are a cripple, what do you want me for?"} P3 recounted receiving backhanded compliments that acknowledged their appearance while simultaneously making inconsiderate remarks about her hearing ability.

\textbf{Misconception.}  Encounters on algorithm-based dating apps sometimes revealed a lack of basic understanding about the DHH community. Participant 3 shared a frustrating experience, \textit{``I encountered someone who assumed that being DHH equates to illiterate, questioning my ability to visit a bookstore and read."} 

P5 often found himself fielding numerous inquiries once he disclosed his DHH identity, 

\begin{quote}
        \emph{``People often pose questions that reveal their complete misunderstanding of DHH communities, such as inquiries about our ability to read or attend school. Although these questions can be misguided, I take the time to respond and educate these hearing people, aiming to enhance public awareness about our community."}
\end{quote} 

\textbf{Lack of Community Support.} A significant challenge was the absence of a supportive and inclusive community on algorithm-based dating apps. Many participants expressed a desire for a dating app that caters specifically to the needs and experiences of DHH individuals. P9 highlighted the difficulty in finding a DHH partner due to the lack of representation on algorithm-based dating apps, \textit{``Accessing a DHH community is essential when looking for a partner who shares my hearing condition. Unfortunately, such a community is scarce on most Chinese dating apps."} P11 sought a more understanding environment by shifting from algorithm-based dating apps to community-based platforms--\textit{Tieba}, where more disabled people gathered, understood, and supported each other. 

\subsection{Interaction Practices}\label{findings: Interaction}
In this part, we explored the communication methods and the strategies they implemented to increase their chances of engaging successfully with potential dates. We further explored certain communication issues experienced by DHH individuals during video calls.


\subsubsection{\textbf{Communication Modalities}} 
\label{4.3.1}
Our participants utilized the rich communication modality features of these online platforms at different times for different purposes with different preferences to facilitate the development of an intimate relationship.

\textbf{Text messages.} Our participants primarily engaged in written correspondence. \textit{Texting} This method was preferred because of its convenience and the comfort it provided to DHH individuals. The asynchronous nature of text messaging offered a buffer during communication, allowing time for thought and reflection, which was particularly valued when communicating with new acquaintances. P12, who used hearing aids, indicated the preference for text messages,
    
    \begin{quote}
        \emph{``Although I can speak and hear with the help of hearing aids, I often find myself reluctant to engage in direct audio conversations with hearing individuals online. It's an area where I'm working to build more confidence."}
    \end{quote}
    
In instances where audio or video communication was necessary, text was used as a supplementary tool to clarify and ensure comprehension. P7, who also utilized hearing aids, combined text with audio calls, \textit{``When clarity becomes an issue during an audio call, I don't hesitate to ask the other person to repeat themselves. At times, we use the chat function to provide written clarifications, which aids in our mutual understanding."} P4 and P11 similarly employed text to support their video conversations, ensuring a seamless exchange of information.

\textbf{Audio messages.} Asynchronous audio messages, however, find utility among the participants, especially when the DHH individuals speak fluently. P9, who was post-lingually deafened and could use spoken language fluently, utilizes a speech conversion tool on his smartphone to interact with voice messages received through social media apps. He shared,  \textit{``My smartphone has a speech converter that I use to translate incoming voice messages to text."} Additionally, he uses voice messaging as a means of providing reassurance to his potential dates, noting, \textit{``I send voice messages only when they express a desire to hear my voice."} 

\textbf{Audio calls.} Real-time audio is occasionally used only when they have a better understanding of each other. P4 and P5, who both used hearing aids and have limited spoken language ability, shared that they would decline audio or video calls from unfamiliar individuals on dating or social platforms. For P9, who did not use hearing aids, found audio features on these Chinese dating apps inaccessible, stating, \textit{``These functions are not practical for my use."}. Conversely, P7 and P8, who used hearing aids but had different levels of fluency in spoken language, reserved audio communication for individuals they have grown more acquainted with. \textit{``Voice or video calls represent a significant level of closeness and are reserved for those I am comfortable with,"} P7 explained.

\textbf{Photos.} \textit{Visual imagery}, particularly photos, emerged as a highly favored form of communication among our participants. P5 and P7 emphasized that exchanging images facilitated a non-verbal means of sharing experiences and emotions, and is a vital aspect of interaction. Similarly, Participants 8, 10, and 12 reported that they frequently share snapshots of their daily lives with prospective dates as a conversational catalyst or to express sentiments non-verbally.

\textbf{Video calls.} Video calls became a preferred method of communication once our participants were ready to take their relationship to the next level. Video calls offered an added layer of verification, allowing participants to compare profile photos with the actual appearance of their counterparts, as noted by P4, P7, P8, and P13. P4 explained a preference for video over audio calls, citing the effectiveness of sign language as a key reason. P13, also stated that he used video calls when he wanted to communicate in sign language. For P12, video calls were invaluable for lip-reading, which was a crucial component of his communication. The use of video calls not only signified a deeper level of intimacy but was also essential—or sometimes the sole option—for DHH individuals who want to advance their relationships. P4 shared that video calls made him feel more connected to his dates, while P12 found video conversations more accessible despite experiencing stuttering. P8 provided insight into the dynamics of video call conversations in the context of online dating, emphasizing the added intimacy and immediacy that this mode of communication affords for DHH people, 
    
    \begin{quote}
        \emph{``In general, conversations with video calls have a higher level of intimacy than those conducted by audio or text. Because you can watch his immediate response. You can change topics right away if you notice that he is having some resistance to this subject. It's important for us. Also, the topics might be more intimate. You could ask him on a video call about why he broke up with his ex-girlfriend. This kind of topic is somewhat personal. When I'm texting, I will avoid asking him questions like this since he has plenty of time to consider them, and I can't see his instant reaction."}
    \end{quote}

\subsubsection{\textbf{Online Community Support}}
\label{section4.3.2}

In this section, we investigated how DHH participants used online community to seek love, including platforms such as \textit{Forums}, \textit{WeChat Public Accounts and WeChat Groups}, and \textit{Human-mediated Matchmaking}. These options exhibited a gradation in the degree of human involvement, with the latter option featuring the most direct human facilitation. Our findings emphasize the benefits of community-oriented approaches in supporting the online dating efforts of DHH individuals in China. This is illustrated by the experiences of our participants: three who are married (P13, P15, and P16) and two in long-term relationships (P9 and P14) found their partners through community-based platforms. Furthermore, many participants expressed more positive experiences with community-based platforms compared to algorithm-based dating apps. These results suggest that community-based platforms better cater to the needs and preferences of the DHH community in online dating in China.

\textbf{Forums.} P11, P14, and P15 used forums to seek romantic connections online and all shared positive experiences.

Participant 15, now married with a child, initially joined \textit{Douban} to find a romantic partner. She actively engaged in posting and participating in various activities on the platform. Eventually, she met her future husband there, who had signed up for \textit{Douban} just to connect with others who shared similar interests. She found this approach to socializing on \textit{Douban} to be more comfortable and observed that the user groups are highly supportive. She also added,
\begin{quote}
    \emph{``\textit{Douban}, due to its larger user base of single young individuals, clear regional divisions, and high activity level, has attracted high-quality users."}
\end{quote}

P14, who also used \textit{Douban} and met her current boyfriend there. She disclosed her DHH identity in the personal profile post, expressed her positive experiences, 
\begin{quote}
    \emph{``The administrators of \textit{Douban} groups are relatively well-regulated. They enforced a policy prohibiting the sharing of personal WeChat contacts within the platform, instead encouraging users to leave their email addresses exclusively within my dating post. This practice necessitates a more formalized method of communication. Consequently, individuals who exhibit interest in my post take the initiative to reach out via email. Most remarkably, the community on \text{Douban} displays a friendly and supportive attitude toward my DHH identity disclosures, with rare instances of derogatory comments or negative responses reported. Instead, some people even expressed encouragement and provided moral support to me. "}
\end{quote}

P11 preferred online communities specifically for disabled individuals. She abandoned algorithm-based dating apps due to unpleasant user experiences and went to another general online forum called \textit{Tieba} ~\cite{tieba}. She mentioned that on algorithm-based dating apps, she seldom initiated conversations and always faced rejection once she disclosed her DHH identity. In the online community for disabled individuals in \textit{Tieba}, she said, 
\begin{quote}
    \emph{``I have posted my detailed introduction, hearing status, and family status on \textit{Tieba} where the communities for disabled individuals could be easily found. If someone is interested in me, he will comment below or send a private message. It is more suitable for a forum with disabled people. Here, we can relate to each other's experiences, unlike mainstream dating apps, where such communities are virtually non-existent. "}
\end{quote}

\textbf{WeChat Public Accounts and WeChat Groups.} Many participants utilized \textit{WeChat}, the most prevalent social app in mainland China, for community-based online love-seeking. WeChat Public accounts and WeChat groups emerged as two popular approaches for finding potential partners within communities.

P5, who preferred a DHH partner, enjoyed the use of WeChat public accounts. He explained,  \textit{``These WeChat public accounts are specifically created for disabled groups, allowing me to connect with individuals who share similar experiences."} P9 and P10 subscribed to these public accounts, which featured profiles with detailed descriptions of disabilities, including DHH status or specific physical disabilities. \textit{``If you are interested in him or her, you can add his or her WeChat private accounts for further communication."} 

Rather than posting personal profiles, our participants primarily used these public accounts to gather information. P10 chose not to post a profile due to the associated costs. Similarly, P12 declined to publish their information on WeChat public accounts.

P10 opted for the WeChat group as it provided a more cost-effective platform to introduce himself compared to the WeChat public account. He shared,
    \begin{quote}
        \emph{``Upon encountering an advertisement on \textit{Douyin}, I decided to join a WeChat group. Within the livestreaming room of a matchmaker, I discovered her private account and added her. She graciously invited me to join a WeChat group consisting of local singles. I prefer WeChat groups due to the presence of matchmakers who offer their assistance."}
    \end{quote}

P8, found the WeChat group to be a valuable space for understanding others' perspectives and getting to know them better. She remarked, \textit{"In the WeChat groups, people engage in discussions about various topics, including matters such as splitting expenses during the initial date. This enables one to identify individuals who share similar viewpoints."} P9 developed an attraction toward a DHH girl he encountered in a DHH WeChat group, 
    \begin{quote}
        \emph{``After joining a WeChat group, I initially observed rather than actively participated. During this time, I paid close attention to the interactions within the group. Eventually, I found a girl who expressed her opinions, which I appreciated. Initially, she declined my friend request on WeChat. However, after I defended her during an argument with another group member, she sent me a friend request."}
    \end{quote}

P16, a married DHH woman who met her husband on a DHH WeChat group, shared her experiences. She actively engaged in discussions related to a television series in the WeChat group. Her participation led to her being noticed by her husband who shared a common interest - her icon picture depicted his beloved anime character. Intrigued by this shared affinity, her husband promptly sent a friend request to her. As their virtual friendship evolved, the two DHH people embarked on a journey of mutual discovery, transcending geographical boundaries to maintain their connection. 

\begin{figure}
    \centering
    \includegraphics[width=0.95\textwidth]{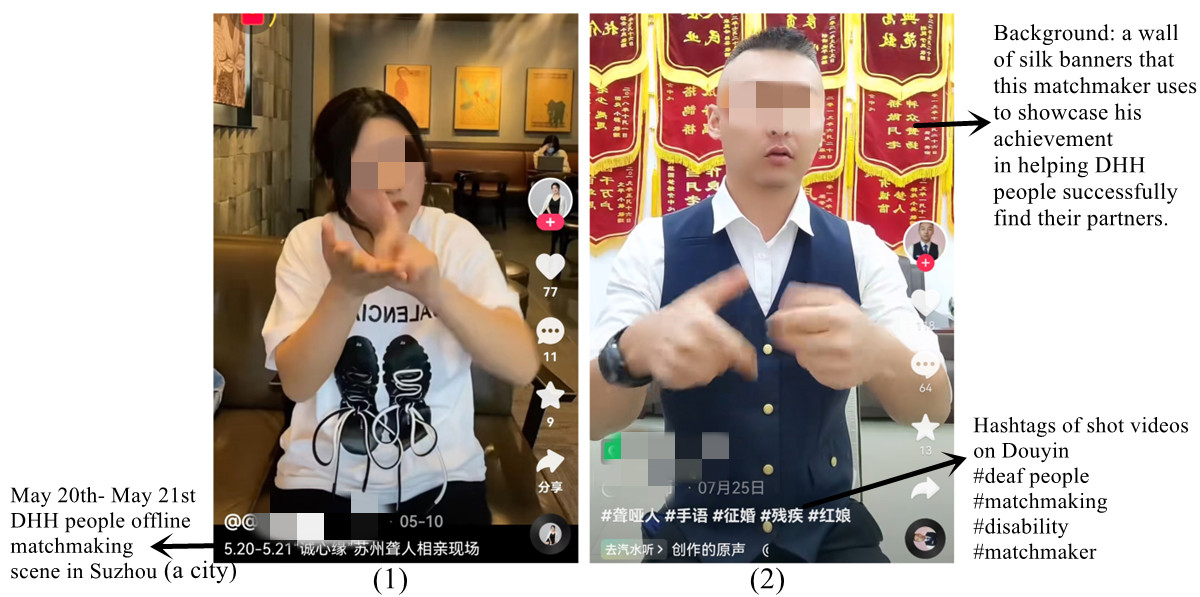}
    \Description[Screenshot of DHH matchmakers' shot videos on \textit{Douyin}]{Two DHH matchmakers are introducing their matchmaking activities and services on \textit{Douyin}. The left picture (1) shows a DHH matchmaker introducing an offline matchmaking activity in sign language. The right one (2) shows a DHH matchmaker introducing his matchmaking services for DHH people with 4 hashtags(\#deaf people, \#matchmaking, \#disability, \#matchmaker) on the screen. He had many silk banners in the background, which means this matchmaker successfully helped many people find their partners. These silk banners were given by his clients as gifts for appreciation. }
   \caption{Two DHH matchmakers are introducing their matchmaking activities and services on \textit{Douyin}. The left picture (1) shows a DHH matchmaker introducing an offline matchmaking activity in sign language. The right one (2) shows a DHH matchmaker introducing his matchmaking services for DHH people with 4 hashtags (\#deaf people, \#matchmaking, \#disability, \#matchmaker) on the screen. He had many silk banners in the background. This indicated the success of the matchmaker who had helped many people to find their partners. These silk banners were given by his clients as gifts for appreciation. }
   \label{fig:DHHMatchmakers}
\end{figure}

\textbf{Human-mediated Matchmaking.} While popular online dating apps like \textit{Tinder} and \textit{Bumble} primarily rely on their own platform algorithms to match compatible individuals, many of our Chinese DHH participants expressed a stronger preference for human-mediated dating or online matchmaking services. In China, algorithm-based dating apps such as \textit{MarryU}~\cite{MarryU} and \textit{Zhenai}~\cite{zhenai} offer additional matchmaking services that require an access fee. According to P10, matchmakers on these apps worked both online and offline, catering to the specific needs of disabled communities. Livestreaming apps such as \textit{Douyin (Chinese Tiktok)} and \textit{Kuaishou}, though not specifically designed for dating purposes, have become popular choices as they attract matchmakers targeting individuals with disabilities, as depicted in Fig. \ref{fig:DHHMatchmakers}. 

P9 described how matchmakers utilize various social media platforms for their matchmaking services, stating, \textit{``These matchmakers have private WeChat accounts on the livestreaming interface or livestreaming profiles. Some specifically focus on disabled groups, where you can find various groups such as blind or visually impaired individuals, DHH individuals, physically disabled individuals, or even individuals with cerebral palsy."}
More than half of our DHH participants expressed the belief that matchmakers can ensure the accuracy of the information provided, particularly regarding physical conditions, and that they have access to ample community resources. P11 elaborated on this process, stating,
    \begin{quote}
        \emph{``The matchmakers undoubtedly conduct a comprehensive evaluation of all the paperwork, including real estate documents, education certificates, ID cards, and disability certifications for individuals like us (DHH). However, many other online dating platforms lack these assurances, especially when it comes to verifying disabilities."}
    \end{quote}
    
P9 mentioned that to participate in an online activity organized by a matchmaker on \textbf{Douyin}, he had to provide his ID card number for identity verification. According to him, \textit{``The matchmakers would verify your identity. Once your verification is successful, you will receive a message confirming your registration. Without this message, you won't be able to attend the activity."} P10 also emphasized the importance of the verification process, especially for online activities organized by offline disabled communities. He explained, \textit{"To register for these events, matchmakers require your ID number and disability certification. Without these documents, attendance is not possible. Therefore, participating in online matchmaking activities within government-certified disabled communities is highly reliable."}

P9 also mentioned a WeChat Public account dedicated to matchmaking for people with disabilities. He said, \textit{"These public accounts have matchmakers who provide assistance. Moreover, they thoroughly review your IDs, disability certifications, and other information before posting your profile on the public account. This ensures that no one can hide their actual physical conditions."}

\textbf{Summary.} The above findings indicate that community-based platforms were superior to algorithm-based dating apps in fostering in-depth communication and increasing the likelihood of successful relationship establishment. This could be attributed to the sense of community and shared values that resonate within common interest groups, which naturally lead to stronger feelings of closeness and trust as relationships progress. As P16 said,
    \begin{quote}
        \emph{``In my DHH WeChat group chats, I find it easier and more reassuring to discuss my interests, as it lacks the drama and distractions often found on other algorithm-based dating apps. Everyone in these groups is a follower of a DHH WeChat public account, creating a sense of closeness and camaraderie among members."}
    \end{quote}
    
Furthermore, for individuals within the Chinese DHH community, the pursuit of authenticity in intimate relationships is paramount, and community-based platforms offer a sense of validation and security that addresses this need for genuineness. Thus, these platforms not only facilitated more profound communication but also provided a safeguard for DHH love seekers, enhancing their online dating experience and potential for finding a compatible partner.

\subsubsection{\textbf{Challenges}}
\label{4.3.3}

Our participants encountered several challenges when seeking love online. These challenges centered on the denominating role of text-based communication environments, difficulties arising from modalities, and the comparatively higher rejection rates they faced on algorithm-based dating apps. 

\textbf{The dual nature of text-based communication.} As previously discussed, our findings indicate that DHH participants heavily relied on text-based communication. This preference suggested that those within the DHH community who were adept at expressing themselves in writing and managing textual information online might have a greater likelihood of finding potential romantic partners. Participants P13, P15, and P16, who all reported having strong written communication skills during their interviews, have successfully married with partners met online. P7, as someone who acquired high writing skills, exemplified someone who had navigated the online dating world with comparative ease. She attributed part of her success to her proficiency in writing: \textit{"I found my three ex-boyfriends online,"} she shared. Therefore, text-based communication offers the opportunity to form more equitable relationships by bridging the gap in spoken communication

However, the heavy reliance on text-based communication may place a high demand on the writing skills of DHH individuals. The absence of vocal tone and inflections in text-based exchanges can make it more challenging to convey and discern subtle emotions and nuances. P13 explained more with an example,

\begin{quote}
    \emph{``A recent news article addressed the tendency of some DHH food delivery drivers to use a direct and unembellished tone in their text messages, a style that might come across as abrupt or impolite to recipients. This communication style is often a consequence of limited exposure to the subtle cues found in spoken language, such as intonation and phonetics, which can lead to a more functional and straightforward expression in written form. Due to their distinct linguistic experiences, DHH individuals may approach texts as a vehicle for clear information, potentially overlooking the emotional subtleties that hearing individuals might expect. In my personal interactions with my wife, I have observed that she displays a level of proficiency in written communication that reflects her understanding of spoken language. The tone and style of her text messages often reveal her linguistic capabilities. "}
\end{quote}

The absence of auditory cues in text communication can have significant implications for Chinese DHH individuals, potentially leading to the impression of inadequate education in the Chinese DHH community. This issue is exacerbated in China, where there is a notable deficiency in specialized training programs for DHH education~\cite{article, inbook}. Consequently, many DHH individuals in China lack adequate education in reading and writing. Those within the DHH community who have underdeveloped language skills are particularly vulnerable to difficulties in navigating text-based communication, which can lead to a sense of exclusion on these online platforms. For example, P8 had a more passive perspective on dating through WeChat groups. She found herself consistently feeling like an outsider in a \textit{WeChat} group created by a friend specifically for dating purposes.
\begin{quote}
    \emph{``The ~\textit{WeChat} group was overwhlemed with text messages. I would receive 999+ messages every day, yet I am really exhausted to read."}
\end{quote}

Therefore, it is essential to understand and investigate the effects of text-based communication on the DHH community in China to foster inclusivity and equality in online spaces. 

\textbf{Audio/Video-Modality Challenges.} Some popular online platforms lacked sufficient assistive tools. P9, for example, used extra hardware or cross-platform software to aid in communication. P6 relied on voice translation provided by her phone. P3 requested a speech-to-text feature for video calls to better comprehend conversations. Since most dating apps don't have sign language transistors, P5 had to use text, \textit{``It's inconvenient because I prefer using sign language. However, if these apps don't have features to support sign language, I always have to resort to text."}, said P5. Although both P13 and P16 often rely on video calls for sign language and lip-reading, they experience inconvenience during poor internet connections. \textit{``Network delays can disrupt video calls, causing the other person's lip movements to be out of sync. This makes lip-reading difficult and interpreting sign language challenging due to the unclear gestures. "} P16 explained. 

Our findings also revealed design elements in some dating apps that inadvertently hinder the use of video calls by DHH users. A case in point is the dating app called \textit{SOUL}, which employs a privacy protection mechanism that superimposes avatars over users' faces during video calls. This feature significantly impedes the ability to lip-read or interpret facial expressions, which are crucial communication tools for the DHH community. P6 expressed frustration with this design, stating, \textit{``This animated cartoon face covers all faces. It was meaningless for our DHH users!"}

\section{DISCUSSION}

Mainstream online dating apps primarily cater to hearing users and have rarely taken into account the specific desires and needs of DHH individuals. This study provides one of the first investigations of the online love-seeking experiences of Chinese DHH users. By conducting semi-structured interviews with 16 DHH participants who have dated online, we found that online dating platforms are appealing to the DHH community due to their large user base, text-based communication options, and visual features that enhance communication effectiveness. We further uncover their dynamic and strategic identity disclosure strategies and how their interactions with the platforms and their users to facilitate love-seeking. In China, while online platforms can help overcome offline dating obstacles, algorithm-based dating apps often fall short in meeting their unique needs, resulting in less favorable experiences compared to community-based platforms. Moreover, these platforms primarily cater to text and spoken communication, offering limited support for sign language users. This lack of support may lead to communication inefficiencies and challenges in self-expression for DHH individuals who rely on sign language. These challenges underscore the pressing need to enhance inclusivity for the Chinese DHH community in online love-seeking environments, and we elaborate on this point below.

\subsection{DHH's Identity Management in Online Love-Seeking}


With the continuous advancement of digital technologies, online love-seeking platforms provide new opportunities for different disabled groups. However, the Chinese DHH community encounters unique and multifaceted identity challenges in online dating. In the following sections, we will explore how people from different disabled groups manage their identity, the dating preference Chinese DHH users, and the design implications for improving their online dating experiences.

\subsubsection{\textbf{Strategic Identity Disclosure}}

Previous studies have shown that individuals with disabilities often prefer to disclose their disability status early. For instance, Porter et al.~\cite{Porter2017} found in their survey of 91 adults with and without disabilities that individuals with disabilities used disclosure as a preemptive filtering method to weed out potentially ableist dating matches. Their findings showed that 57\% of disabled respondents preferred to disclose their disabilities early in interactions to avoid wasting time and emotional investment on potential partners who might react negatively. Additionally, Martino and Kinitz ~\cite{martino2022s} examined adults with intellectual disabilities, finding 62\% favored early disclosure to mitigate negative reactions and improve their chances of finding compatible partners.

Our research sheds new light on the online love-seeking behaviors of Chinese DHH individuals, revealing a more flexible approach to identity disclosure than previously documented. Unlike earlier studies that mostly focused on early disclosure, our findings indicate that Chinese DHH participants make disclosure decisions based on the depth of their conversations and the specific context (section.\ref{Section4.2.2}). Some participants choose to disclose their status early, but many adapt their timing according to the situation and the traits of their dating partners. Additionally, the timing of disclosure is even adjusted to match each partner’s characteristics and the level of interest. This flexible and nuanced strategy allows DHH individuals to better manage potential stigma and rejection. By tailoring their disclosure approach to the dynamics of each interaction, DHH individuals may enhance their dating experiences. Our findings suggest that these adaptive strategies are important in understanding how DHH individuals navigate online dating.



Another distinct feature of the DHH community is that the disclosure is closely tied to the mode of communication. In those initial online contacts, communication via texts would make their disabilities ``invisible", like what we found in section.\ref{4.3.1}. The visibility of the DHH status depends on when synchronized audio is needed, as disabilities can be ``invisible to certain perceivers under certain conditions"~\cite{davis2005invisible}. Previous studies suggests that that ``visible" disabilities are more likely to be disclosed early compared to ``invisible" disabilities~\cite{Porter2017}. In our study, participants agreed that they would disclose their hearing status because it will eventually become apparent during a audio/video meeting or in-person interaction. 

Future work might explore how various communication technologies either support or hinder the identity disclosure of different disabled groups. Evaluating how specific features on online dating platforms influence the timing and comfort of identity disclosure for disabled groups could be helpful. Also, it may be insightful to investigate the impact of real-time communication tools, such as video or audio calls, livestreaming, and live-chatting room on disclosure decisions. Research could also focus on the effect of varying levels of anonymity provided by different platforms on the willingness of disabled group members to share personal information. Finally, gathering user feedback to enhance technological features that facilitate sensitive disclosures might improve user experiences for the disabled groups.

\subsubsection{\textbf{Situating DHH love-seeking within Social and Cultural Context}}

Many of our study participants showed a tendency to date hearing individuals, as indicated in Table \ref{table:participants}. This preference is behind the desire to expand their social circles through the use of mainstream online dating platforms. This inclination towards dating hearing individuals may also suggest a limited sense of self-identification within a DHH-centric community. A key contributing factor is the notable fragmentation within the Chinese DHH community, where individuals use various regional sign languages and spoken languages, leading to communication challenges. This stands in contrast to Western contexts where a dominant sign language like American Sign Language (ASL) serves to unify DHH communities at a national level. While the specific reasons behind the partner preferences of DHH users are not within the scope of our study, it is essential to recognize the potential of online dating platforms in addressing the needs of Chinese DHH users. Additionally, cultural influences such as traditional matchmaking practices may impact our participants' utilization of dating platforms, as discussed in Section \ref{section4.3.2}.  Our participants are accustomed to human-mediated matchmaking, possibly influenced by traditional matchmaking culture in China~\cite{He2023}.

The distinct online dating motivates and experiences of DHH individuals in China underscore the significance of considering diverse social and cultural backgrounds to enrich our comprehension of online dating practices. Exploring how local dating norms, relationship expectations, and communication preferences influence disclosure decisions among participants in different regions in future work could offer valuable insights. The development of online dating platforms that are attuned to social and cultural nuances could improve user satisfaction. By contextualizing love-seeking within local cultural landscapes, our goal is to advance knowledge and enhance support for online dating experiences among minority groups, addressing their distinct challenges and preferences.

In conclusion, there is significant potential for online platforms, particularly algorithm-based dating apps, to enhance their design by incorporating features that cater to the linguistic and cultural intricacies of the Chinese DHH community. In the following sections, we will explore specific design recommendations aimed at reducing obstacles and promoting a more inclusive online environment for individuals seeking love. 


\subsubsection{\textbf{Design implications}}

\begin{enumerate}
    \item \textbf{Design for Navigating DHH Identity on algorithm-based dating apps.} 

     Algorithmic constraints can limit the visibility of content produced by disabled groups, such as Blind or Low Vision (BLV) streamers, as algorithms may not recognize their content as potentially popular~\cite{Rong2022}. This bias reflects a broader issue of imprecise algorithmic filtering, which can restrict diversity and impede meaningful connections. In the context of dating apps, these algorithmic limitations are also evident.  Users are required to create profiles that balance authenticity and attractiveness, often relying on static images and brief descriptions~\cite{Ward2017,MacLeod2019}.  However, algorithms designed for the majority group tend to filter matches based on limited criteria, overlooking personal characteristics that could impact compatibility~\cite{Chan2021}.  Our study found that personality embedded in the text messages plays a crucial role in successful matching for DHH individuals. However, algorithms often overlook this nuanced aspect, which can be relatively more important for minority groups, resulting in less successful and less inclusive platforms that do not cater well to DHH individuals..

    In addition, in our interviews, some participants were more open to dating hearing individuals while others expressed a preference for connecting. However, current algorithm-based dating apps do not offer the necessary features to assist DHH users in easily finding and connecting with their preferred community. As a result, some DHH users turn to community-based platforms where they can rely on matchmakers or intermediaries to facilitate introductions. 

    For our DHH participants, the effectiveness of the recommendation algorithms may be further worsened by the lack of accessible features such as video introductions or enhanced text communication. The text-based communication prevalent on dating apps, which is the primary interaction method on most dating platforms (Section.\ref{4.3.1}), can be inefficient and restrictive when sign language is not accommodated.

    To create a more inclusive environment for DHH users, algorithm-based dating app developers should consider implementing features like video introductions, enhanced text communication, and more transparent, customizable algorithms. Additionally, it is important to establish reliable methods for verifying disability status within these platforms. This approach could potentially allow for more rigorous verification of information, ensuring the authenticity of potential partners' disability status, which may foster trust and facilitate meaningful connections. However, we must be cautious, as such features could inadvertently contribute to further segregation between DHH and hearing users in the pursuit of romantic connections. We encourage future researchers to examine this issue before widespread implementation to mitigate potential divisions.

    \item \textbf{Design for Addressing Bias and Discrimination.} Another challenge impacting the DHH community on online platforms is the handling of bias and discrimination. Our participants frequently encountered instances of hostility or discrimination when revealing their identity, yet hesitated to utilize the available reporting functions. On the one hand, discrimination often takes the form of subtle rejection, where many contacts stopped responding or gave various excuses to end the conversation once they was told about the participants' DHH status. These rejections and implicit discrimination have led to a withdrawal from algorithm-based platforms for our DHH participants. Additionally, the reluctance to report such behaviors is often due to platforms relying on telephone follow-ups, which do not cater to the communication needs of DHH users.

    To address these challenges and reduce discrimination, online platforms should implement more robust reporting systems that are easy to use and accessible for DHH individuals. Hutson et al.~\cite{Hutson2018} emphasized the importance of designing technical systems to be resistant to bias and discrimination. Specifically, they noted that platforms with comprehensive and user-friendly reporting tools were more effective in mitigating bias and providing safer environments for marginalized users. Implementing text-based live chats, in-app messaging systems with customer support, or video calls with sign language support can serve as good alternatives to telephone-based reporting mechanisms, making it easier for DHH users to report issues effectively and comfortably.

    Moreover, platforms should include educational resources and training for all users to foster a more inclusive environment. By encouraging education and voluntary affirmation of inclusive pledges, platforms can prompt users to reflect on their own preferences and attitudes. Hutson et al.~\cite{Hutson2018} discussed how educational initiatives can signal the importance of understanding and openness as platform norms. These initiatives can visibly mark the profiles of users who promise to uphold these norms, allowing users to signal both their intimate preferences and broader social attitudes. This approach not only promotes inclusivity but also reduces the stigma and discrimination faced by marginalized groups.

    Furthermore, utilizing AI-driven moderation tools to identify and mitigate DHH-oriented discriminatory behavior in real-time is worth-trying. As a new trend to mitigate online offensive messages~\cite{Schulenberg2023}, AI-driven moderation may monitor interactions for signs of hostility or bias, addressing issues before they escalate. For DHH users, these tools could be beneficial as they can provide a safer and more responsive online environment. AI moderation could also detect subtle forms of discrimination that might be missed by human moderators~\cite{Schulenberg2023}, ensuring that DHH users do not have to continuously report issues themselves, thereby reducing the emotional and cognitive load on these users.

    These design implications aim to create a more inclusive and supportive online environment for DHH individuals, hopefully reducing the impact of discrimination and enhancing their ability to engage meaningfully on online love-seeking platforms.

\end{enumerate}

\subsection{DHH's Communication Modalities in Online Love-Seeking}

Our research offers valuable insights into how DHH individuals utilize different communication modalities in online dating. By extending Media Richness Theory (MRT)~\cite{Daft1986} to consider the specific needs and preferences of DHH users, we can better understand their choices and behaviors in this context. Furthermore, our findings suggest that the concept of media richness needs to be redefined and extended in the context of DHH individuals' use of online dating platforms.

\subsubsection{\textbf{DHH Communication Modalities in Online Love-Seeking through the lens of Media Richness Theory}}

Our research offers valuable insights into how DHH individuals utilize different communication methods in seeking love online, a perspective that can be analyzed through the framework of Media Richness Theory (MRT)~\cite{Daft1986}. According to MRT, various communication media have different capabilities in conveying information. Rich media, like face-to-face communication, are effective for complex and nuanced interactions as they can provide multiple information cues simultaneously. In contrast, researchers noted that in general settings, leaner media, such as text messages, are more suitable for straightforward information exchange~\cite{Daft1986,Dennis1999,Ashley2022}.

In section.\ref{4.3.1}, we observed that DHH individuals have preferences for different communication modalities at various stages of intimate relationships. Text messaging, classified as a lean medium, is commonly used during initial contacts because of its restricted information transmission. This limitation in information exchange serves as a much-needed buffer in communication between DHH individuals and strangers from online dating platforms. However, when seeking to establish a deeper connection, DHH individuals found video chat to be more effective for in-depth conversations. Video chat, categorized as a rich medium, allows for visual cues and sign language, facilitating the expression of complex information and emotions, thereby enhancing mutual understanding and supporting the growth of romantic relationships. Audio calls, positioned as an intermediate medium between rich and lean, are the least favored due to their synchronized nature, which poses significant communication challenges for DHH individuals. This observation extends the application of MRT to the context of DHH individuals seeking love, highlighting the importance of utilizing different communication modalities at different stages of relationship development. For DHH individuals using love-seeking platforms, it is not always true that rich communication media are always preferred.

We have discussed the communication modalities of DHH users through the framework of Media Richness Theory, not only confirming but also extending aspects of the theory to encompass the adaptive strategies and unique needs of DHH individuals. For DHH users, the definitions of lean and rich media need to be reconsidered. Visual media like videos and video calls, which are typically considered rich media, may require high fidelity of facial features to allow lip-reading for this media to be truly rich for DHH individuals. On the other hand, combining text (captions) and video in the right way can enhance media richness for DHH users, even though MRT would not necessarily suggest this in the context of hearing individuals. Moreover, the limited reading and writing abilities of some DHH individuals can impact the richness and usability of text-based media, which are usually classified as lean media.

\subsubsection{\textbf{Multi-modal Design on Algorithm-based Dating Apps}}

Participants in our study expressed a strong preference for more visual information to improve communication on dating platforms, especially for DHH individuals who rely on lip-reading ~\ref{4.3.1}. However, some algorithm-based dating apps use avatars that obscure facial features and are not trained to replicate lip movements, which can significantly reduce the richness of this visual medium for DHH users. We propose implementing advanced facial capture technology in avatars to accurately depict lip movements and facial expressions in real time. This idea is supported by existing research, such as the system outlined by Philipp, which combines KinectV2 input for skeletal tracking and an RGB camera for facial data, enabling the real-time animation of a photorealistic 3D avatar model ~\cite{7532378}.

Our participants prefer video chats as their interactions progress because it allows them to see their potential partners' physical appearance and nonverbal cues. However, the current speech-to-text services are limited to voice messages in text chat and are not available during video calls in most Chinese dating apps. This limitation reduces the richness of video chat for DHH users who rely on captions to fully understand the conversation. To address this issue, we recommend incorporating real-time captioning and sign language interpretation directly within online dating platforms. Traditional captioning method can be challenging for DHH users as they need to quickly match captions with visual content, leading to a disjointed experience~\cite{1035891}. Therefore, we propose implementing innovative captioning solutions that overlay captions directly onto the video feed in various positions. This dynamic captioning approach could improve the viewing experience for DHH users by simplifying the process of connecting spoken words with visual context, thereby increasing the richness of the video medium.

Finally, as online love-seeking for DHH individuals heavily relies on text-based communication, there is a concern that those with limited writing and reading proficiency may encounter challenges and potential marginalization within this context. For these individuals, the lean nature of text-based media is further exacerbated, potentially hindering their ability to effectively communicate and form connections online. Therefore, it is worth considering the exploration of tools specifically designed to aid DHH individuals in adapting to text-based environments. This represents a promising avenue for future research that could improve inclusivity and accessibility within the online dating landscape.

\subsection{The Importance of Community Support in Online Love-Seeking}


Community support is instrumental in cultivating positive online network relationships across diverse domains such as education~\cite{Gautam2021,Wilner2024}, employment, and workspace~\cite{Hui2014,Groot2024}. Gautam and Rosson~\cite{Gautam2021} emphasize how community infrastructures can enhance students' sense of connection and improve their learning outcomes. Groot et al.~\cite{Groot2024} explore the significance of community in fostering professional relationships and teamwork in remote settings. In the realm of love-seeking, community support is also vital. Masden and Edwards ~\cite{Masden2015} found that external forums linked to online dating sites create a supportive “outsourced community” where users share experiences, offer advice, and support each other. Moreover, livestreamer-moderated matchmaking communities were shown to provide substantial support for single seniors~\cite{He2023}. Match-seekers could freely communicate, initiate relationships, and receive emotional support from peers in matchmaker-managed groups, which helped reduce loneliness and foster a sense of community. Shen et al. ~\cite{Shen2024} highlight the importance of community support in voice-based dating platforms like Soul, where various chat room formats foster genuine emotional connections and personality expression.

We had similar findings in our study (Section.\ref{section4.3.2}), indicating that community-based platforms offer more positive online dating experiences. Despite not being originally tailored for dating, these platforms create a supportive environment for DHH individuals. By providing an online space for DHH individuals to connect and receive enhanced community support, these platforms enable more comfortable discussions around emotional and romantic topics. This supportive setting alleviates the challenges and stress related to disclosing disabilities, empowering users to have greater control over their expectations compared to mainstream platforms.

Moreover, our research indicates that community support is not only based on the DHH identity. A broader disabled community can also provide significant support. Additionally, more personalized and intellectually stimulating online platforms, such as \textit{Douban}, known for mutual support and a welcoming atmosphere, can foster intimate relationships.  Therefore, our research suggests that while shared identity within a community is relevant,the crucial factors for a community to be friendly and advantageous for online dating are its inclusive and open-minded approach, rational and objective discussions, and closer interpersonal connections~\cite{He2023,Hutson2018}. The development of community-centric online dating platforms with user-friendly interfaces and features that promote safe and comfortable disclosure is essential. Future research may explore how communities can enhance the online dating experiences of minority groups. For example, longitudinal studies can assess how online community dynamics can impact relationship building. Furthermore, examining the effectiveness of online community regulatory policies in reducing disability-related stigma would be beneficial. By emphasizing the role of community support, future research can expand on our findings and contribute to creating a more supportive and inclusive online dating environment for various minority groups

\section{LIMITATION AND FUTURE RESEARCH DIRECTION}

This study has several limitations. We discuss these limitations and what future work can be done to address these limiations.

\subsection{Participant Recruitment and Community Representation}

The recruitment method involved sourcing participants from the social media platform \textit{Xiaohongshu} and snowball sampling. This method have attracted a young cohort and those not in a long-term relationships. Moreover, as Table.\ref{table:participants} shows, only two out of sixteen participants were users of sign language, reflecting a potential under-representation of deaf individuals who primarily communicate using sign language. This findings from this selected group may not represent the full spectrum of experiences within the Chinese DHH community. 


\subsection{Positionality and Inclusion of DHH Researchers}
None of the authors in this study are DHH individuals. This absence may have introduced unintended biases and limited our ability to fully understand and interpret the experiences of the DHH participants. The power dynamics between hearing and DHH individuals could influence the research outcomes despite our best efforts to conduct thorough and unbiased research.

\subsection{Recommendations for Future Research}
Given these limitations, future research should strive to include a more diverse array of DHH participants, especially those who are profoundly deaf and primarily use sign language. Additionally, exploring how online dating platforms can integrate more inclusive communication tools, such as sign language support and visual aids, would be imperative to genuinely enhance the dating experience for all DHH individuals. Furthermore, caution must be exercised when extrapolating these results to the entire DHH population, especially to older age groups. This highlights the need for broader research to capture the full diversity of the DHH community's online love-seeking experiences.

Future studies should also consider broader recruitment strategies to include participants from more established phases of life, ensuring that the insights and recommendations cover a wider range of the DHH population. Moreover, future research directions should prioritize the inclusion of DHH researchers in the research team. By fostering a more inclusive and participatory research environment, we aim to bridge the gap between hearing and DHH perspectives, ensuring that the research findings are more representative and impactful. Collaborative efforts with DHH researchers can lead to the development of more effective and culturally appropriate solutions tailored to the needs of the DHH community.


\section{CONCLUSION}
This study investigates how Chinese DHH individuals navigate empowerment, identity, and interaction dynamics, as well as the barriers they encounter while seeking love on mainstream Chinese online love-seeking platforms. By interviewing 16 Chinese DHH users, we gained insights into how algorithm-based dating apps and community-oriented platforms make initial interactions and relationships more accessible by reducing the need for auditory communication compared to offline settings. Our finding further highlights the significance of community support and the various approaches and timing participants use to disclose their DHH identity. Additionally, we identified three primary challenges in identity management for DHH participants: experiences of social exclusion, addressing misunderstandings, and feeling disconnected from DHH communities.


In terms of interaction practices, our participants primarily used text-based messages, audio transcriptions, images, and occasionally videos. There was more positive feedback from community-based platforms like \textit{Forums}, \textit{WeChat Public Accounts and WeChat Groups}, and \textit{Human-mediated Matchmaking} compared to algorithm-based dating apps. Drawing from our analysis, considering linguistic and cultural factors, we have proposed incorporating multi-modal design features and creating a matching system tailored to the needs of the DHH community to improve the inclusivity of online platforms. We hope this research will encourage further exploration of digital inclusivity for marginalized groups and that our recommendations will aid in the design of algorithm-based dating apps, benefiting the DHH community in their pursuit of love.

In conclusion, our study acknowledges the challenges faced by Chinese DHH individuals in their quest for companionship on Chinese online platforms. The research underscores the marginalization and biases encountered by DHH participants when revealing their identities, prompting some to avoid dating apps due to feelings of exclusion. However, Chinese DHH individuals often find more welcoming environments and positive interactions in community-based platforms, despite not being primarily designed for dating. While community-based platforms showed more positive experiences for love-seeking, our findings highlight the necessity for algorithm-based dating apps to consider social factors like deaf culture awareness and sign language diversity.


The difficulty in finding romantic partners online reflects a significant societal gap, influenced by limited understanding of the Chinese DHH community. Bridging this gap goes beyond technology alone. We advocate for the future development of dating apps in China and globally that incorporate a deeper understanding of social factors. By creating a social environment that appreciate a diversity of social, physical, and personal identities, we can reduce social barriers within and between DHH communities and the broader society. Inclusive platforms have the potential to foster more meaningful connections, breaking down barriers and promoting a society that values diversity in all relationship forms. This call to action aims to inspire innovation in online dating platform design, offering improved opportunities for connection that embrace and accommodate the diverse human experience.

\bibliographystyle{ACM-Reference-Format}
\bibliography{main}

\appendix
\section{Questions for Preliminary Survey and Semi-structured Interview}
\label{Questions}

\subsection{Preliminary Survey}

\begin{enumerate}
  \item What is your age?
  \item What is your sex?
  \item What is your education level?
  \item What is your DHH state (shown in DHH certificate)?
  \item What is your communication state (both hearing and speaking)?
  \item Which platforms do you use to seek romantic partners online?
  \item How long have you used online love-seeking platforms? Frequency?
  \item Have you tried to attend matchmaking activities online?
\end{enumerate}

\subsection{Semi-structure Interview}

\textbf{Interaction (RQ1\&RQ2):}

\begin{enumerate}
  \item What motivates you to seek love online?
  \item What are the biggest differences between your online dating experience and offline dating?
  \item Are there any aspects of online dating that you find better than offline dating? Or worse?
  \item How do you attract other love-seekers online? 
  \item Have you ever disclosed your DHH identity while seeking love online? If so, in what way and where did you disclose your identity?
  \item (If disclosed) What kinds of replies you received? And how you replied back? What were the feelings?
  \item What is the most memorable experience you've had with online dating? Or perhaps something that made you happiest or unhappiest? Would you mind sharing? Did these events happen before or after you disclosed your identity?

\end{enumerate}

\textbf{Relationship (RQ2):}

\begin{enumerate}
  \item What is your love-seeking state?
  \item Do you have any people or stories happened online you want to share with me?
  \item How do you meet him/her? which platform and when?
  \item How do you interact with him/her online?
  \item Do you build connections with him/her offline? How?
  \item Do the interaction approaches change as time goes by? Online or Offline? How?
  \item What is your preference for your potential partners' hearing state?
  \item Are there any other stories in online love-seeking that you want to share with me?
  
\end{enumerate}

\textbf{Platforms and Challenges (RQ3):}

\begin{enumerate}
    \item (If multiple platforms were used) Do you use the same or different approach across different platforms? Why?
    \item (If multiple platforms were used) What kind of replies you received? Are they the same or different across different platforms? What do you think could be the reason behind this phenomenon?
    \item Which platforms are more convenient or more difficult to use? Why?
    \item Which design do you like or dislike in current platforms?
    \item Have you encountered any specific difficulties of online love-seeking?
\end{enumerate}

\end{document}